\documentclass[12pt]{article}

\usepackage{amssymb,amsmath,amsfonts,eurosym,geometry,ulem,graphicx,caption,color,setspace,sectsty,comment,footmisc,caption,pdflscape,subfigure,array,hyperref}
\usepackage[affil-it]{authblk}
\usepackage[
backend=biber,
style=alphabetic,
sorting=ynt
]{biblatex}

\begin{document}

\title{The Limits of AI in Financial Services}

\author[1]{Isabella Loaiza}
\author[1]{Roberto Rigobon}
\affil[1]{Sloan School of Management, Massachusetts Institute of Technology}

\date{}
\maketitle

\begin{abstract}
AI is transforming industries, raising concerns about job displacement and decision-making reliability. AI, as a universal approximation function, excels in data-driven tasks but struggles with small datasets, subjective probabilities, and contexts requiring human judgment, relationships, and ethics. The EPOCH framework highlights five irreplaceable human capabilities: Empathy, Presence, Opinion, Creativity, and Hope. These attributes are vital in financial services for trust, inclusion, innovation, and consumer experience. Although AI improves efficiency in risk management and compliance, it will not eliminate jobs but redefine them, similar to how ATMs reshaped bank tellers' roles. The challenge is ensuring professionals adapt, leveraging AI’s strengths while preserving essential human expertise.
\noindent \\
\vspace{0in}\\
\noindent\textbf{Keywords:} AI, Financial Services, Human Capabilities\\
\vspace{0in}
\end{abstract}

\setcounter{page}{0}


\section{Main} \label{sec:introduction}

Almost every sector faces disruption from the rise of Artificial Intelligence, driven by two core fears. The first is substitution—the concern over which jobs will remain once machines and algorithms can perform most tasks. The second is AI’s judgment—an existential fear reminiscent of The Matrix, where an advanced AI might determine that humans are more harmful than beneficial to global progress. These anxieties have shaped public debate across industries, and the financial system is no exception.

In our paper “The EPOCH of AI” by Loaiza and Rigobon \cite{loaiza2024epoch}, we take a different approach to the question of labor market substitution, and implicitly, also dispel the fear of AI’s judgement. This approach – as we highlight in the paper – is complementary to a very large existing literature, but it is distinct in one important dimension: instead of analyzing which tasks and jobs the current technology can substitute, we ask what capabilities will AI never be able to substitute.

Artificial Intelligence is fundamentally built on universal approximation functions—an idea that mathematicians and engineers have studied for centuries. The origins of function approximation trace back to Euclid and Archimedes (c. 300 BCE), who used geometric methods to estimate areas and volumes. In the 14th century, Madhava of Sangamagrama (India) developed power series expansions, laying the groundwork for what would later become the Taylor series. Then, in the 1660s, Isaac Newton introduced polynomial approximations and the binomial expansion, providing the foundation for Brook Taylor’s formalization of the Taylor series in 1715. The goal was always the same: to approximate complex functions with simpler ones. The 18th and 19th centuries saw major advancements in approximation theory, including Fourier series for periodic functions, Chebyshev polynomials, and Carl Friedrich Gauss’s least squares approximation. 

These methods broke down complex functions into more manageable components, forming the backbone of numerical analysis. Today, modern AI models—from Random Forests, to Neural Networks to Large Language Models—are all forms of universal approximators. They take highly intricate processes and approximate them using vast collections of simpler functions, enabling machines to learn, predict, and generate with remarkable accuracy.

As highlighted by Loaiza and Rigobon \cite{loaiza2024epoch}, universal approximation functions have inherent limitations when: The dataset is small; the model must extrapolate far beyond the training data; the process allows for multiple acceptable outcomes; the desired outcome is a relationship rather than a recommendation or action; the process relies on subjective probabilities rather than realized probabilities.

Although these challenges are technical, they align with core human abilities. We refer to these as human capabilities, encapsulated in the EPOCH framework:
\begin{itemize}
    \item Empathy and Emotional Intelligence
    \item Presence, Networking, and Connectedness
    \item Opinion, Judgment, and Ethics
    \item Creativity and Imagination
    \item Hope, Vision, and Leadership
\end{itemize}

In the financial system, many critical aspects of financial services are deeply reliant on EPOCH, highlighting the irreplaceable role of human expertise in areas where AI falls short.

This is not the first time a technology threatens the financial system – nor will it be the last. In fact, the introduction of ATMs is one of the most talked about examples in every single business school to highlight how technology does not destroy jobs, it can also enhance them. Before ATMs, bank tellers’ jobs centered around money exchange with customers – either to accept deposits or withdrawing them. The fear was that ATM’s were going to substitute bank tellers. In practice, ATM’s substituted the most time-consuming activity the tellers were doing, freeing them to be able to perform other services. Instead of declining, the number of bank tellers in the US soared. The key for the bank tellers was to invest in tasks and education on complementary activities to the technology – instead of investing time and effort in those aspects that competed with it.

Therefore, the question was, what can the teller do that is complementary to the ATM? That is exactly our approach, what the financial sector can do that is complementary to the AI tools that are likely to appear in the future?

Financial services offer many benefits. Some are clearly AI driven.

\subsection{Benefits and Challenges of AI in Financial Services}

\textbf{Liquidity provision.} Liquidity provision ensures that funds are readily available for transactions, lending, and investment without causing market disruptions. AI can enhance this process by predicting liquidity needs, optimizing capital allocation, and automating market-making activities. Machine learning models analyze vast datasets in real time to forecast liquidity shortages, adjust asset allocations, and detect potential liquidity crises before they unfold. High-frequency trading algorithms and AI-driven market makers efficiently provide liquidity by continuously adjusting buy and sell orders, reducing spreads, and stabilizing financial markets.

\textbf{Risk Management.} AI-driven models assess credit risk, detect fraud, and manage market fluctuations. AI can significantly enhance this activity by analyzing large datasets in real time to detect emerging risks, predicting market trends, and automating decision-making processes. Machine learning models can evaluate creditworthiness, monitor transactions for signs of fraud, and assess portfolio risk by simulating various market scenarios.\\

\textbf{Portfolio Investment.} The financial system already offers robo-advisors! The greatest advantage is to be able to provide real-time data analysis that improves aspects such as tax harvesting and other services.\\

\textbf{Efficiency.} For example, algorithmic trading involves using automated, pre-programmed strategies to execute trades in financial markets at high speeds and volumes. AI significantly enhances trading by detecting patterns, and identifying trading opportunities faster than human traders.\\

\textbf{Regulation.} Almost every bank today uses machine learning tools to detect money laundering. Compliance with regulation has been already enhanced by AI - detecting hidden patterns in the data has been and will be a game changer for compliance.

These are all activities where AI has and will continue to have major impact. These tasks are clearly at risk, but the most fundamental question is, are the jobs at risk? Or will these jobs be enhanced as it was the case with bank tellers?

There are other aspects in financial services that cannot be substituted by machines - Activities and tasks that are fundamental to the provision of the service. This is the topic we would like to spend most of our time on.\\

\textbf{Trust}: According to Psychologists trust is the belief in the reliability, integrity, and ability of another person or system. It is the foundation behind one person's feelings of confidence in another one, and it is the fundamental building block of cooperation and connection. It requires repeated positive experiences, consistency, transparency, and quite importantly, the emotional safety that the trusted party will not exploit the other vulnerabilities. In finance, trust is the most important attribute any financial institution has. The act of delegating the investment of deposits and savings to an institution is the ultimate act of trust – and it is conceivable to argue that every single financial crisis, in the end, involves a violation of trust. An AI agent can have positive experiences repeated in a consistent manner, as well as a certain amount of transparency – although interpretable machine learning algorithms are very rarely as good as the non-interpretable ones. But it is the confidence that the information will not be used against the user that constitutes a major blockage for AI to fully develop trust. Most people argue that trust is about a correct answer, and as the psychologist's definition highlights, it is not about the answer, but about the process.\\

\textbf{Financial Inclusion.} Financial inclusion is a virtue – in moral philosophy terms. It plays a crucial role in promoting economic equality and development. It refers to the process of ensuring that individuals and businesses, particularly those from underserved or disadvantaged communities, have access to affordable and appropriate financial services. This is an example in which AI algorithms fail terribly. The reason is that there is very little data about those individuals that have been excluded from the financial system, and furthermore, it relies on insufficient statistics.

An example on accepting a student to MIT might illustrate the point. Imagine there are two highschoolers that have GPA’s of 4.2 and 4.1. In principle, 4.2 seems higher than 4.1. However, the hiring of a worker is the search for talent, not grades. So, let us put into context each grade. The 4.2 is a person that comes from a very good family, lived in a very good neighborhood, went to private school and did summer camps in Harvard. The The second one comes from a single mother, who lived in inner Chicago. The mother is working two jobs to be able to put her kids in school, he went to public school, and he worked on the weekends to complement the household income. Who is more talented?

Of course the advocates of AI would say that if we had all the information about the second kid the AI would be able to make a better decision. But that is incorrect. That information does not exist to the extent that can be used by an universal approximation function. The reason is that there are not that many examples. So, ultimately, it is a matter of judgement. We humans, at least some of us, believe in the virtuosity of equality of opportunities. Therefore, it is within that context that we are willing to make a decision in favor of someone that we believe has faced a more challenging situation, even though the formal statistic suggests the opposite. In other words, our subjective beliefs are in conflict with the estimated probabilities. It has been proven that machines are fraught with many biases (racism, sexism just to name a few), and they can even be more biased  than humans. The reason is that the computer instead of judging the situation is more likely to act on “statistics” – even quite incomplete ones – instead of unique context. This is an example where the AI agent faces two of its biggest challenges - small data sets, and the a disagreement between the subjective probabilities and the realized ones.\\

\textbf{Innovation.} Innovation is a tricky concept. Innovation can involve transforming creative concepts into practical solutions that add value and improve efficiency. From that point of view, the process of innovation is a trial and error procedure. An AI agent can do that perfectly, and conceivably faster than a human. After all, playing board games is an act of innovation within that definition – and we all know how good computers have become at board games.

But innovation is also about having the ability to produce something new, something that is far from the data we have observed, something that might have never existed before. This is the case of Art and Music. We asked Google who were the most innovative painters and why? The top five were Da Vinci, Van Gogh, Monet, Picasso, and Dali. The reason why is because they all dared to challenge the common wisdom of their time. 

From the statistical point of view, what made them great was the ability to extrapolate far from the space where other works of art existed at that time. Innovation and experimentation have a purpose and a degree of randomness at the same time. Universal approximation functions have a chaotic behavior when extrapolating far from the observed data, and that behavior could become innovation, but also it is mostly randomness without purpose. 

For the record, the evaluation and ranking of the painters is not because we are good at art. We used AI tools to do so. This is a task where AI is wonderful – summarization of the opinions of experts. Google answered with Gemini, their generative AI. The actual quote we obtained is the following: “they significantly pushed the boundaries of art through their unique styles, experimentation with techniques, and exploration of new concepts, often influencing subsequent art movements.” This is the summary of the views of hundreds of critics and art experts that describe the greatness of these artists. Gemini is good for the summary but might not be that good for the judgement.\\

\textbf{Consumer Experience.} Finally, financial services are services. Trust is developed through a repeated interaction, but also by human connection. It is not the same to ask a chat box than to ask a person. And we know most people will say that is not the case, but it is. We use the following example as evidence that only in small stakes do we trust a chat box. 

Imagine that you type the following in Google:
\textit{“I have a lot of pain in my left arm, with swelling, and abnormal growth in the bone. Furthermore, it breaks very easily and does not heal well, what condition could I have?”} 

Imagine that the answer is the following:
\textit{“You have a rare form of bone cancer that typically starts in the arm and is often described as 'eating the bones' – the technical term is osteosarcoma. Osteosarcoma is considered a highly aggressive cancer that can rapidly spread to other parts of the body, and you should amputate your arm.”} 

The question is how many would immediately amputate their arm without any further consultation? \\

Most would go to the doctor (human) take 100 tests, and when the doctor says: \textit{“You have a rare from…”} (exactly the same as Google) most will not stop there. Most will seek a second opinion. Which will not be asking ChatGPT what to do. They will seek another human. And after another set of tests, if the answer continues to be that the cancer is too aggressive and it cannot be treated, and that the only solution is to amputate the arm, then, and only then, most will proceed to remove it. 

In this example, Google was right the first time. However, trust is not about the answer is about the process. Consumer financial services are not about the answer, it is about the procedure. It is about the relationship that humans develop by the fact that we are present.
 
For the record, Google did not recommend amputation (although that is one of the available treatments). Google exactly said:
\textit{“If you are experiencing significant pain in your left arm with swelling, abnormal bone growth, easy fractures, and poor healing, you likely have a condition called bone tumor or a related issue like osteoporosis with pathological fractures; it is crucial to consult a doctor immediately for diagnosis and treatment as this could be a serious medical concern.”} In other words, Google recommended asking a human. 

Financial services are very likely to experience a massive shock from Artificial Intelligence. But the shock will not be about jobs disappearing, but about jobs being reinvented and redesigned. Financial services are fundamentally related to innovation, service, inclusion and trust. Those are virtues that are rooted in human capabilities. If there is a challenge at hand, however, it will be about understanding how we can transform current jobs to complement the technology fast enough to make the transition as smooth as possible.

\singlespacing

\clearpage

\onehalfspacing

\clearpage

\clearpage
\printbibliography


\end{document}